\documentclass[preprint,8pt]{elsarticle}

\usepackage{amssymb}
\usepackage{amsmath}
\usepackage{multirow}
\usepackage{booktabs} 
\usepackage{graphicx}
\usepackage{hyperref}
\usepackage{pifont}
\newcommand{\xmark}{\ding{55}}%
\usepackage{color}
\newcommand*{\Scale}[2][4]{\scalebox{#1}{$#2$}}%
\newcommand{\model}{GraphRevisedIE}%

\usepackage{lineno}

\journal{Pattern Recognition}

\begin{document}

\begin{frontmatter}



\title{GraphRevisedIE: Multimodal Information Extraction with Graph-Revised Network}


\author[1]{Panfeng Cao}
\author[2]{Jian Wu}

\affiliation[1]{organization={University of Michigan},
            city={Ann Arbor},
            postcode={48109}, 
            state={MI},
            country={USA}}
            
\affiliation[2]{organization={University of Science and Technology of China},
            city={Hefei},
            postcode={230026}, 
            state={Anhui},
            country={PR China}}            

\begin{abstract}
Key information extraction (KIE) from visually rich documents (VRD) has been a challenging task in document intelligence because of not only the complicated and diverse layouts of VRD that make the model hard to generalize but also the lack of methods to exploit the multimodal features in VRD. In this paper, we propose a light-weight model named \model{} that effectively embeds multimodal features such as textual, visual, and layout features from VRD and leverages graph revision and graph convolution to enrich the multimodal embedding with global context. Extensive experiments on multiple real-world datasets show that \model{} generalizes to documents of varied layouts and achieves comparable or better performance compared to previous KIE methods. We also publish a business license dataset that contains both real-life and synthesized documents to facilitate research of document KIE.
\end{abstract}




\end{frontmatter}


\section{Introduction}
\begin{figure*}
    \centering
    \includegraphics[scale=0.15]{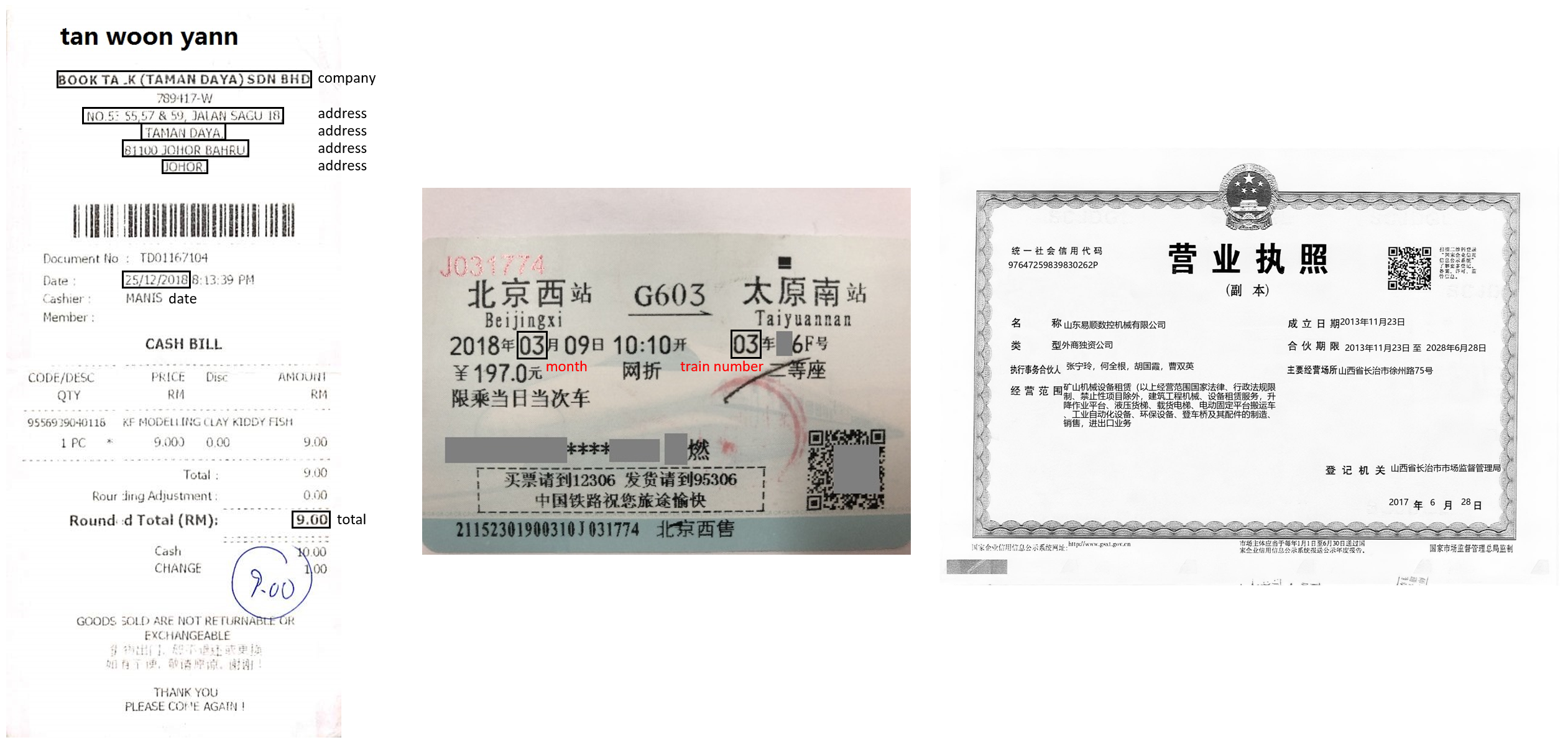}
    \caption{Example VRD of different layouts. (a) Key entities to be extracted are marked with red rectangles. (b) Same text 03 results in semantic ambiguities for different entities. (c) Example business license. }
    \label{fig:example}
\end{figure*}

Optical character recognition (OCR) is a technology to recognize the texts in the scanned documents, and KIE is the downstream task of OCR that extracts entity information from the texts. KIE is critical to applications such as document indexing, information archival, and information retrieval \citep{JUNG2004977} because it can save significant amounts of time and resources. Deep learning based approaches have become the focus of modern research and achieved state-of-the-art (SOTA) results. However, it remains a challenge to effectively utilize the multimodal features in visually rich documents (VRD). As we can see in Figure \ref{fig:example}, VRD can be a structured or unstructured document such as a receipt, ticket, business license, etc. There are varied formats, layouts, and contents in VRD, and multimodal information is critical to resolving the semantic ambiguity, which can occur when textual information alone is not enough to distinguish the entities. For example, in Figure \ref{fig:example}(b), texts of both the month and train number are \texttt{03}, but they are of different entity types. We can only distinguish them from the layout and visual information.

Traditionally, template or rule based KIE methods \citep{smartFIX,6628593} have been widely adopted in commercial applications. The interpretability of those methods makes the KIE program easy to be adopted and scaled for different scenarios. However, significant engineering efforts and domain-specific knowledge are needed to design the handcrafted rules and patterns for different entities. And those methods only support a limited number of document types and cannot deal with complicated and unstructured documents. Later KIE systems formalize the problem as a Named Entity Recognition (NER) task, which typically applies models to predict the beginning-inside-outside (BIO) tags of the tokens. \citep{DBLP:journals/corr/MaH16, DBLP:journals/corr/ChiuN15, TOLEDO201927} are based on bidirectional LSTM (BiLSTM) and add an additional convolutional neural network (CNN) layer to enrich the feature representation for entity extraction. \citep{DBLP:journals/corr/abs-1809-08799} is a pure image based approach that introduces a fully convolutional encoder-decoder network based on the VGG architecture. The semantic features are encoded with the document layout to perform information extraction. Other methods \citep{DBLP:journals/corr/abs-1708-03743, DBLP:journals/corr/abs-1808-09101} use graph based long short-term memory (LSTM), which allows a varied number of incoming edges at each memory cell, to jointly learn the entities and relations extraction. While those methods have been proven effective, they do not make full use of the multiple modal features available in the VRD and cannot tackle semantic ambiguity.

Efficiently combining multimodal features in the VRD has become the focus of modern research. Graph based methods represent the document as a graph, with nodes representing segments and edges representing segment relations. Graph convolution is utilized to propagate the global context and enrich the feature embeddings. \citep{DBLP:journals/corr/abs-1810-13083, DBLP:journals/corr/abs-1903-11279} utilize predefined graphs to combine the textual and layout features. \citep{DBLP:journals/corr/abs-2004-07464} uses a graph learning convolutional network (GLCN) \citep{8953909} to dynamically learn the graph and generate a richer semantic representation of the segment. Lately, pre-training based methods such as LayoutLMv2 \citep{DBLP:journals/corr/abs-2012-14740}, StrucText \citep{DBLP:journals/corr/abs-2108-02923} and BROS \citep{DBLP:journals/corr/abs-2108-04539} have been proposed to deeply fuse multimodal features from large-scale pre-training datasets and achieved SOTA performance in downstream KIE tasks. However, those models have a higher number of parameters and require larger datasets for effective pre-training. Furthermore, it's also difficult to deploy and maintain the models in real-life settings due to the complicated multi-stage training paradigm.

In this paper, we propose a novel, lightweight framework named \model{} to tackle the problem of multimodal feature embedding. Textual, visual, and layout features are jointly embedded in the model to address the semantic ambiguity. We design a graph module inspired by \citep{DBLP:journals/corr/abs-1911-07123} to learn the graph representation for the document by graph revision and perform graph convolution to enrich the multimodal feature embedding with global context. The graph module also leverages the sparsification technique to learn the appropriate graph representation for the sparse document.

The main contributions of this paper are summarized as follows:
\begin{itemize}
\item In this paper, a novel framework named \model{} is proposed to handle document KIE. Multimodal features in VRD are effectively embedded to cope with the semantic ambiguity.
\item As far as our knowledge, \model{} is the first graph based model that utilizes the graph revision technique in document KIE. The graph module can effectively learn document graphs and contextualize the multimodal feature embedding with global context.
\item We publish a dataset that contains real-life and synthesized business licenses to facilitate the document KIE research. 
\item Extensive experiments on multiple public datasets show that \model{} outperforms existing graph based models. The model also has comparable performance to pretrained models, while it has significantly fewer parameters and does not depend on large pre-training datasets~\footnote{Our code and business license dataset are publicly available at \url{https://github.com/AYSP/GraphRevisedIE}.}.
\end{itemize}

\section{Related Works}
Early research on entity extraction focuses on the exploitation of a single modal feature from the document. \citep{DBLP:journals/corr/LampleBSKD16} utilizes BiLSTM to embed the textual feature. \citep{DBLP:journals/corr/abs-1708-03743, DBLP:journals/corr/abs-1808-09101} introduce the graph based LSTM that supports cross sentence semantic relation and entity extraction. \citep{DBLP:journals/corr/MaH16, DBLP:journals/corr/ChiuN15} utilize both LSTM and CNN to get better textual embedding to perform sequence labeling. \citep{gui-etal-2019-lexicon} leverages plain-text semantic features from the document but does not exploit the layout and image features. \citep{DBLP:journals/corr/abs-1809-08799, DBLP:journals/corr/abs-1909-09380} use the image features to encode the semantic contents but leave the textual and layout features untapped. \citep{DBLP:journals/corr/abs-1810-13083, DBLP:journals/corr/abs-1903-11279} apply BiLSTM to embed the textual features and GCN to incorporate the layout features. Graph convolution is used not only to propagate the global contexts but also to generate the node embedding. \citep{DBLP:journals/corr/abs-1810-13083} depends on task-specific graphs and needs predefined data structures, which are hard to extend to other types of documents. \citep{DBLP:journals/corr/abs-2004-07464} jointly embeds the textual and visual features and uses the absolute layout feature in the graph module to generate the node and edge embedding. 

Pretrained transformer encoder based approaches achieved SOTA results by deeply fusing multimodal features in pretraining. LayoutLM \citep{DBLP:journals/corr/abs-1912-13318} designs pre-training tasks that utilize the absolute 2D layout features and the textual features. Visual features are embedded in fine tuning. LayoutLMv2 \citep{DBLP:journals/corr/abs-2012-14740} moves the visual feature embedding to pre-training and learns effective multimodal feature representation. However, those approaches depend on a large corpus and need significantly more parameters and time to train. 

Compared to previous graph based methods \citep{DBLP:journals/corr/abs-1903-11279, DBLP:journals/corr/abs-2004-07464}, \model{} differs in several aspects. First, \citep{DBLP:journals/corr/abs-1903-11279, DBLP:journals/corr/abs-2004-07464} choose fully connected graph as the initial graph. Although they can dynamically update the edge weights during training, they do not support adding new edges due to the element-wise product in the attention function. If the edge weight is learned to be 0, it is removed and cannot be added back. When the document graph is highly sparse, the model eventually learns a suboptimal graph representation. Nevertheless, the graph module in our framework does not enforce a fully connected graph as the initial graph and it supports adding new edges as well as updating existing edge weights. Attention based graph convolution is performed to contextualize feature embedding with global context to facilitate final prediction. Furthermore, our graph module does not require a loss function and is seamlessly integrated with the downstream learning objective, which implicitly entices the graph representation learning. Finally, we rely on relative positional information to embed the layout feature instead of using absolute positional information, which can introduce spatial bias in the case of image twisting, shifting, and rotation. Relative positional information better captures the global invariant layout relations of entities and helps improve the model's performance. 

\section{Model Architecture}

\begin{table}[]
\centering
\resizebox{0.73\columnwidth}{!}{%
\begin{tabular}{ll}
Symbol      & \multicolumn{1}{c}{Meaning}                               \\
\toprule
$D$         & The document                                              \\
$I$         & The scanned document image of $D$                         \\
$T$         & Text segments recognized by OCR in $D$                    \\
$N$         & Number of text segments in $D$                            \\
$L$         & Maximum text segment length of $T$                        \\
$t_{i}$     & $i_{th}$ text segment                                     \\
$c_{j}^{i}$ & $j_{th}$ character of $t_{i}$                             \\
$B$         & Bounding boxes of $T$                                     \\
$b_{i}$     & $i_{th}$ bounding box                                    \\
$d$         & Model dimension                                           \\
$A$         & Weighted adjacency matrix representing the graph          \\
$S$         & Similarity matrix of nodes in the graph                   \\
$v_{i}$     & $i_{th}$ node corresponding to $t_{i}$                    \\
$a_{ij}$    & Weight of the directed edge from $v_{j}$ to $v_{i}$       \\
$d^{n}$     & Embedding dimension in the graph module                   \\
$TE$        & Textual embedding of $T$                                  \\
$te^{i}$    & Textual embedding of $t_{i}$                              \\
$VE$        & Visual embedding for segments corresponding to $B$        \\
$ve^{i}$    & Visual embedding for the segment corresponding to $b_{i}$ \\
$d^{b}$     & Sinusoidal embedding dimension                            \\
$PE_{ij}$   & Relative positional embedding of $b_{i}$ and $b_{j}$      \\
$SE$        & Segment embedding for segments corresponding to $B$       \\
$HE$        & Hidden embedding of $SE$                                  \\
$DE$        & Document embedding of $D$                                 \\
$d_{tags}$  & Number of predefined BIO tags                             \\
$Z_{ij}$    & Probability of the $i_{th}$ character token being the $j_{th}$ tag \\
$T_{ij}$    & Transition probability from the $i_{th}$ tag to the $j_{th}$ tag \\
$y_{i}$     & $i_{th}$ predefined BIO tag                               \\
$Y_{DE}$    & The set of all possible tag sequences for $DE$            \\
\bottomrule
\end{tabular}%
}
\caption{Notations.}
\label{tab:notations}
\end{table}

\begin{figure*}
    \centering
    \includegraphics[scale=0.20]{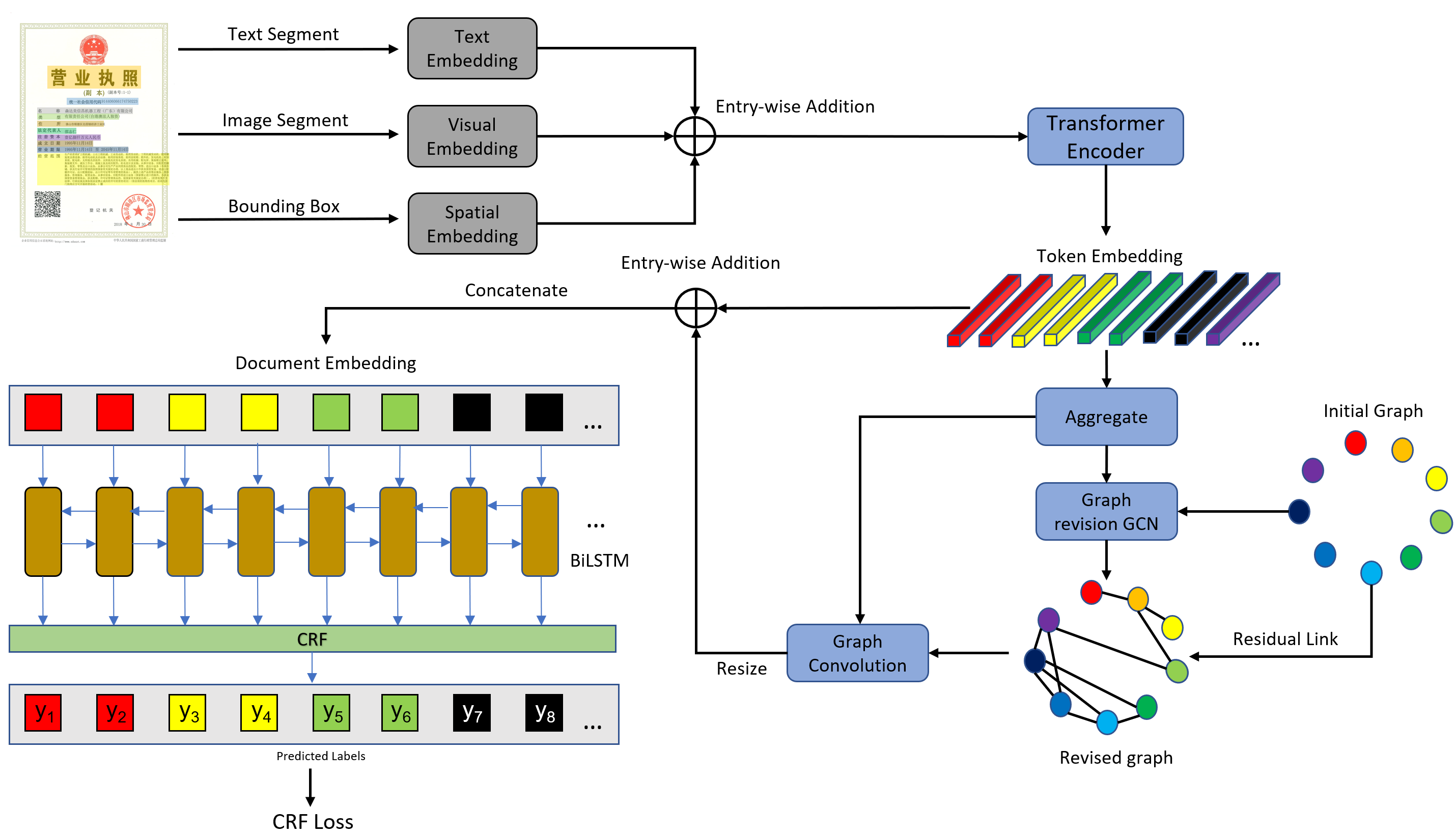}
    \caption{Overall diagram of the GraphRevisedIE framework. Note that for illustration purposes, we use the same color for all tokens in the same segment and different colors for tokens in different segments. The top section of the diagram demonstrates the process of multimodal feature fusion. The bottom right section explains the graph module for feature embedding enrichment. Self-connected edges are omitted. The bottom left section is the BiLSTM-CRF module that calculates the CRF loss and produces the final prediction.}
    \label{fig:model-arch}
\end{figure*}

Table \ref{tab:notations} gives the notations used in the paper. Given $D$ and $I$, we first use an open source OCR tool (e.g. Tesseract~\footnote{\url{https://tesseract-ocr.github.io/}}) to recognize $N$ segments that correspond to the nodes in the graph. The graph is represented by the weighted adjacency matrix $A$, in which the element is the edge weight. The model architecture is illustrated in Figure \ref{fig:model-arch}, which comprises three modules: a multimodal feature embedding module, a graph module, and a decoding module. 

\subsection{Embedding}
As shown in Figure \ref{fig:model-arch}, the multimodal feature embedding module has three branches, each embedding a single modal feature. First, for textual embedding $TE$, it includes all segment textual embeddings: 

\begin{equation}
TE = Concat(te^1, ..., te^N) \in \mathbb{R}^{N \times L \times d} 
\end{equation}
\begin{equation}
te^i = Concat(Emb(c^i_1), ..., Emb(c^i_L)) \in \mathbb{R}^{L \times d}
\end{equation},
where $Concat$ is the concatenation operation and $Emb: \mathbb{R} \rightarrow\mathbb{R}^{d}$ is the character token embedding function, e.g. one-hot embedding.

\begin{figure*}
    \centering
    \includegraphics[scale=0.15]{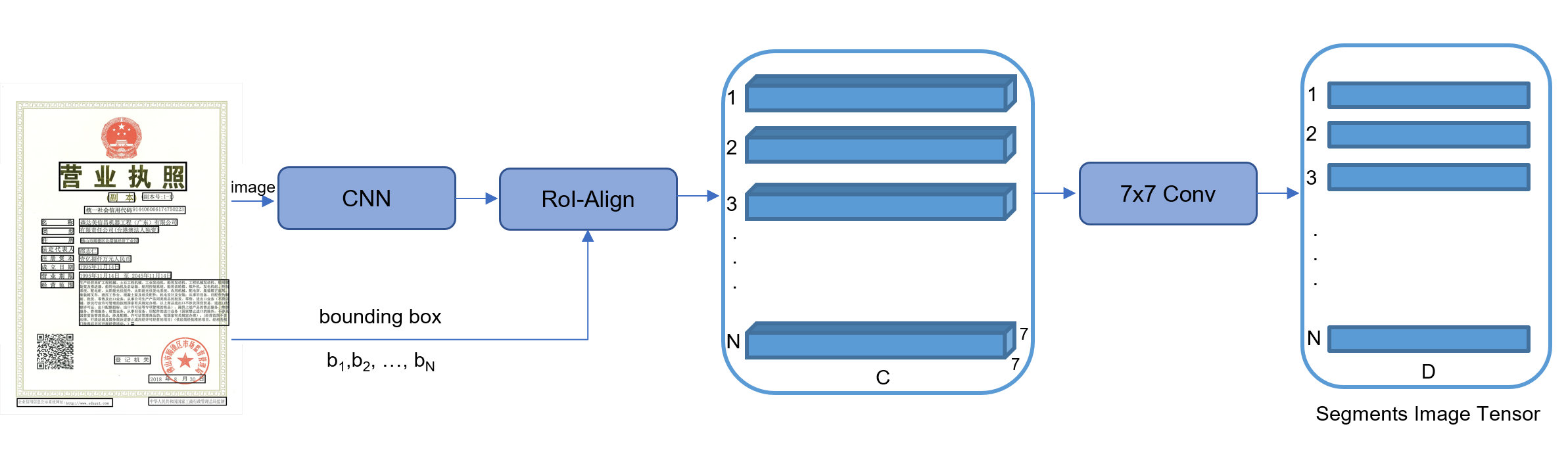}
    \caption{Illustration of generating the image embedding. Inputs are the raw image and bounding boxes of segments. RoI-Align is used to extract segment level features from the whole image feature produced by the CNN module. A convolution kernel is applied to transform the output dimension of RoI-Align to the model dimension.}
    \label{fig:visual-embedding}
\end{figure*}

Then we use a CNN as the visual feature extractor to get the visual embedding. The visual features of the segment, such as font, size, and color, can help enrich the segment embedding. As presented in Figure \ref{fig:visual-embedding}, a CNN module is first used to get the global feature maps of the whole image, and then the local feature map of each bounding box is extracted from the global feature maps via RoIAlign\citep{DBLP:journals/corr/HeGDG17}. Finally, we apply the convolution on the local feature map to generate the segment level visual embedding $ve^i$. Given $I$ and $B$, $VE$ is calculated as follows:
\begin{equation}
VE = Concat(ve^1, ..., ve^N) \in \mathbb{R}^{N \times d}
\end{equation}
\begin{equation}
ve^i = Conv(RoIAlign(CNN(I), b_i)) \in \mathbb{R}^{d}
\end{equation}
Within $b_i$, all characters share the same $ve^i$ by design. The final form of $VE$ after resizing is:
\begin{equation}
VE = Concat(ve^1, ..., ve^N)\in \mathbb{R}^{N \times L \times d}
\end{equation}
\begin{figure*}
    \centering
    \includegraphics[scale=0.25]{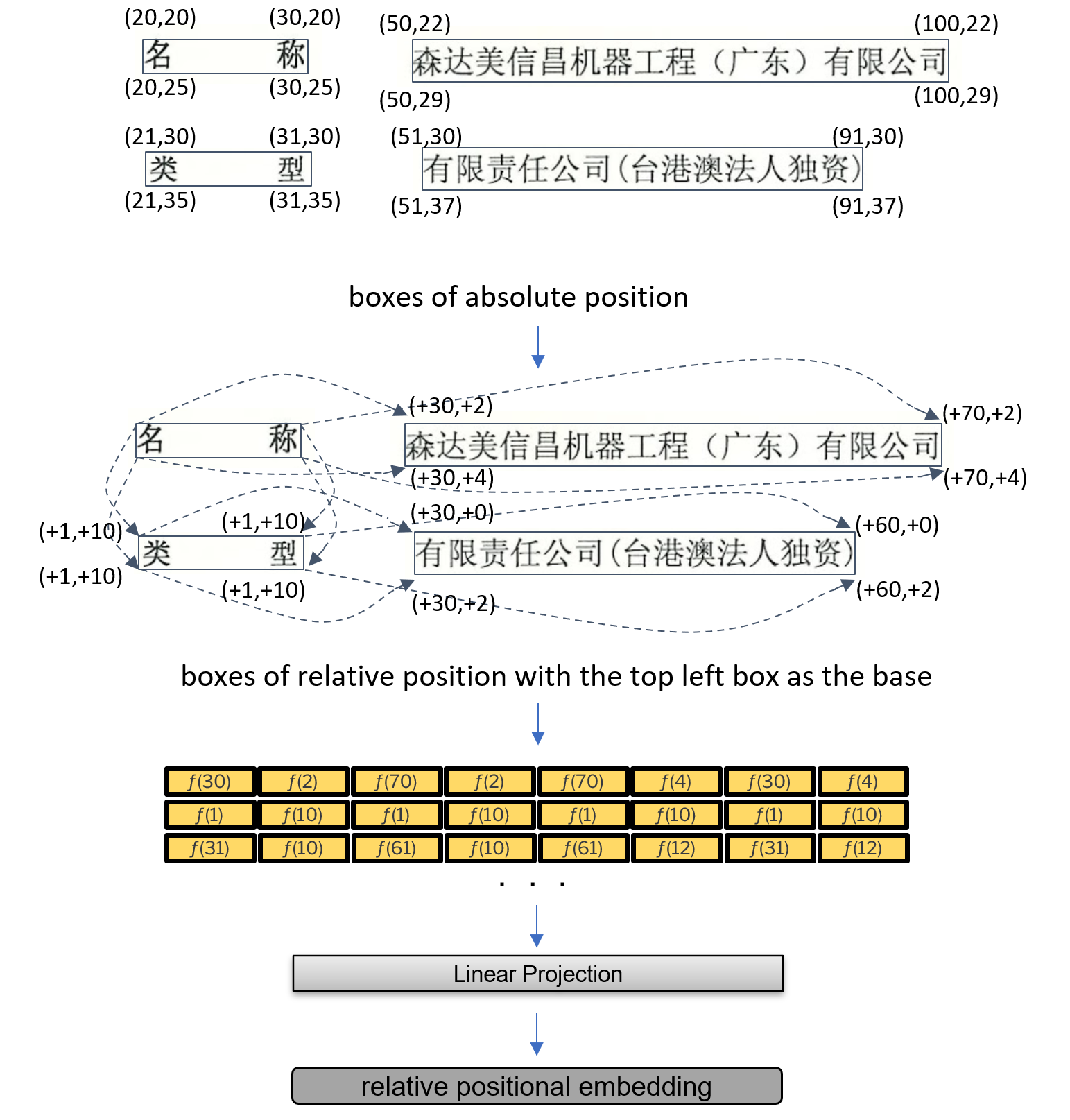}
    \caption{Process of generating the relative positional embedding. Relative positions are first embeded with the sinusoidal embedding function $f$ and then go through a linear projection layer to get the final embedding.}
    \label{fig:relative-pe}
\end{figure*}

Finally, we embed the layout features. Inspired by the 1D positional embedding in Transformer \citep{DBLP:journals/corr/VaswaniSPUJGKP17}, we designed the 2D relative positional embedding, which normalizes the spatial relations between segments. It’s robust to the positional shifting caused by the raw image distortion and helps the model learn the inherent layout. Given $T$ and $B$, we first normalize the coordinates so they fall between 0 and 100. Then for $t_{i}$, $b_i$ and $t_{j}$, $b_j$, we calculate the relative positional embedding between those two bounding boxes in the following equation:
\begin{equation}
\begin{split}
PE_{ij} = Concat( & f^{sinu}(x^{tl}_i - x^{tl}_j), f^{sinu}(y^{tl}_i - y^{tl}_j), \\
& f^{sinu}(x^{tr}_i - x^{tr}_j), f^{sinu}(y^{tr}_i - y^{tr}_j), \\
& f^{sinu}(x^{br}_i - x^{br}_j), f^{sinu}(y^{br}_i - y^{br}_j), \\
& f^{sinu}(x^{bl}_i - x^{bl}_j), f^{sinu}(y^{bl}_i - y^{bl}_j)) \cdot W \\
& \in \mathbb{R}^{d}
\end{split}
\end{equation},
where $f^{sinu}: \mathbb{R} \rightarrow\mathbb{R}^{d^b}$ is the sinusoidal embedding, which is used in \citep{DBLP:journals/corr/VaswaniSPUJGKP17} to help embed the relative positions of segments. $W: \mathbb{R}^{8 \times d^b} \rightarrow \mathbb{R}^d$ is the linear projection matrix that maps from the sinusoidal embedding dimension to the model dimension. The process is also illustrated in Figure \ref{fig:relative-pe}, where we use the top left box as the base box for comparison. It is worth noting that we embed all four vertices of the text segment, allowing the projection matrix to learn spatial features such as relative height, width, and distance. Since tokens in the same segment share the same bounding box coordinate, we resize and get the final relative positional embedding $PE \in \mathbb{R}^{N \times L \times d}$. 

By now we have all the single modal embeddings, we calculate the merged multimodal embedding by performing element-wise addition of those embeddings and applying transformer encoding:
\begin{equation}
E = transformer\_encoder(TE + VE + PE) \in \mathbb{R}^{N \times L \times d}
\end{equation}
\subsection{Graph Module}

\begin{figure*}
    \centering
    \includegraphics[scale=0.22]{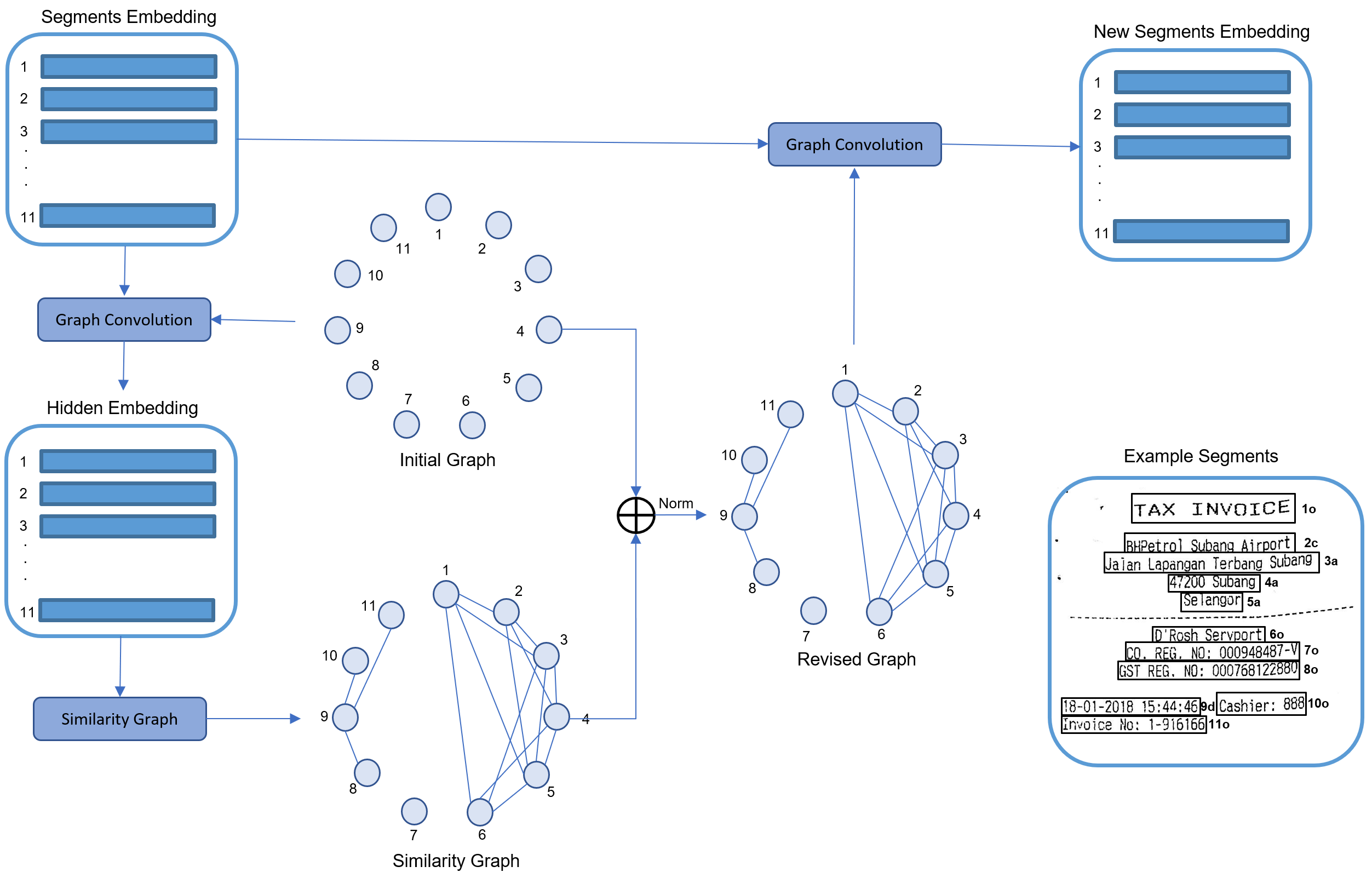}
    \caption{The graph module illustrated on an example SROIE receipt. In the bottom right, segments corresponding to the nodes are given with the indexes and labels (o: other, c: company, a: address, d: date). We use an identity matrix as the initial graph. For simplicity, self-connected edges are omitted. A new segment embedding is produced by graph convolution on the revised graph using the original segment embedding.}
    \label{fig:graph-module}
\end{figure*}

The graph revised module propagates the non-local and non-sequential contexts among segments to enrich the segment embedding. Although similar, our graph module design differs from \citep{DBLP:journals/corr/abs-1911-07123}, which constructs a single large graph for the entire dataset to address the node classification problem. For the document KIE task, our graph module needs to learn the graph representation for all documents in the dataset. The initial graph of each document is represented by an identity matrix, and the graph module revises it to find the appropriate graph.

As is described in Figure \ref{fig:graph-module}, we perform two operations in this module, i.e. graph revision and attention based graph convolution. Given $D$, we first aggregate the character multimodal embeddings in the segments to produce the segment embedding $SE \in \mathbb{R}^{N \times d}$. With the initial weighted adjacency matrix $A$, we calculate the hidden embedding $HE$ from $SE$:
\begin{equation}
HE = A \cdot \tanh(A \cdot SE \cdot W_1) \cdot W_2,
\end{equation}, where $A \in \mathbb{R}^{N \times N}, W_1 \in \mathbb{R}^{d \times d^n}, W_2 \in \mathbb{R}^{d^n \times d}, SE, HE \in \mathbb{R}^{N \times d}$ and $\tanh$ is the activation function. 

The similarity matrix $S$ of segments is then derived from $HE$: 
\begin{equation}
S = Knn(Kernal(HE, HE^T)) \in \mathbb{R}^{N \times N},
\end{equation}We use dot product as the kernel function following \citep{DBLP:journals/corr/abs-1911-07123}. Since $S$ is dense, the K nearest neighbor ($Knn$) algorithm is applied to sparsify the graph and only keep the top K neighbors of each node. Unlike \citep{DBLP:journals/corr/abs-1911-07123}, where sparsification is mainly for memory and computation efficiency since important neighbor nodes are relatively constant, this sparsification process is necessary in our task because the entity in the document can be split into an unknown number of continuous multi-line segments. $Knn$ helps identify important neighbors efficiently by removing unimportant edges. To obtain the revised adjacency matrix $A'$, we add $S$ and $A$ and normalize the result: 
 \begin{equation}
A' = Norm(A + S) \in \mathbb{R}^{N \times N} 
 \end{equation}With the element-wise addition operator $+$, new edges can be added and existing edges can be reweighted.

Using attention-based graph convolution with $A'$, we compute the updated segment embedding $SE'$: 
\begin{equation}
SE' = A' \cdot SE \cdot W_{3} \in \mathbb{R}^{N \times d}, W_{3} \in \mathbb{R}^{d \times d}
\end{equation}Compared with GLCN \citep{8953909}, our graph module does not have the graph representation learning loss, which simplifies the design and experiments. Note that although the graph module helps the model generalize on documents with varied and complex layouts, it is not indispensable when the document has a relatively fixed layout. We study the importance of the graph module in ablation study (\S \ref{sec:ab}) on various document datasets. 

\subsection{Decoding}
We resize the segment embedding output by the graph module to $N \times L \times d$ and add it to the multimodal character embedding to get the final embedding. As a result, the character embedding combines not only the textual, image, and layout features of its own segment but also the global context of neighboring segments. 
To begin decoding, we concatenate all character embeddings in the segments from left to right and from top to bottom to produce the document level embedding $DE \in \mathbb{R}^{[N \cdot L] \times d}$. The reason we do the concatenation is because if a sentence is broken down into several text segments, we can restore its original semantic structure. The document embedding series is passed to a BiLSTM model to encode the long short-term dependencies, and the prediction scores of BIO tags are calculated by:
\begin{equation}
Z = BiLSTM(DE) \cdot W_B \in \mathbb{R}^{[N \cdot L] \times d_{tags}}
\end{equation}, where $W_B \in \mathbb{R}^{d^b \times d_{tags}}$ is the linear projection matrix mapping from the hidden dimension of BiLSTM to the output BIO tags dimension. $Z$ is the scores matrix, in which $Z_{ij}$ indicates the possibility of $i_{th}$ token being the $j_{th}$ tag. 
Finally, character level BIO tagging is performed via a CRF layer. CRF is particularly effective in NER tasks where token labels have strong interdependencies. The tagging decisions for the tokens are jointly considered for the document series. Given a sequence of predictions $y = (y_1, y_2, ..., y_n)$, the score is defined as: 
\begin{equation}
s(DE, y) = \Sigma^{N \cdot L}_{i=0} T_{y_i, y_{i+1}} + \Sigma^{N \cdot L}_{i=1}Z_{i, y_i}
\end{equation}, where $T$ is the transition matrix of scores. $T_{y_i, y_{i+1}}$ is the score of transitioning from $y_i$ to $y_{i+1}$. $y_0$ is the start tag and $y_{N \cdot L + 1}$ is the end tag. The conditional probability of $y$ given $DE$ is calculated with the softmax operation: 
\begin{equation}
p(y|DE) = \frac{e^{s(DE, y)}}{\Sigma_{\bar{y}\in Y_{DE}}e^{s(DE, \bar{y})}}
\end{equation} The loss function is the logarithm of the conditional probability:
\begin{equation}
Loss = -\ln(p(y|DE)) = -s(DE, y) + \underset{\bar{y} \in Y_{DE}}{lnadd}\hspace{0.1cm} s(DE, \bar{y})
\end{equation}The optimal tag sequence is the one with the highest conditional probability: 
\begin{equation}
y^* = \operatorname*{arg\,max}_{y \in Y_{DE}} p(y|DE)
\end{equation}We search the optimal tag sequence with dynamic programming. 
\section{Experiments}
\subsection{Datasets}
Our model is evaluated on multiple real world public datasets: SROIE \citep{DBLP:journals/corr/abs-2103-10213}, CORD \citep{Park2019CORDAC}, FUNSD \citep{jaume2019funsd}, Train Tickets\citep{DBLP:journals/corr/abs-1909-09380} and Business Licenses.

\begin{table}[]
\centering
\Scale[0.8]{
\resizebox{\columnwidth}{!}{%
\begin{tabular}{|l|l|l|l|l|}
\hline
Dataset          & Type    & Language & \# Keys & \# Images                    \\ \hline
SROIE            & Receipt & English & 4 & Train 526, Val 100, Test 347 \\ \hline
CORD             & Receipt & English & 30 & Train 800, Val 100, Test 100 \\ \hline
FUNSD           & Form     & English & 4   & Train 149, Val 0, Test 50 \\ \hline 
Train Ticket     & Ticket  & Chinese & 8 & Train 1749, Val 100, Test 80 \\ \hline
Business License & License & Chinese & 9 & Train 1120, Val 100, Test 100 \\ \hline
\end{tabular}
}}
\caption{Statistics of each dataset.}
\end{table}

\textbf{SROIE} dataset is used to extract entity information from scanned receipts. It contains 626 receipts for training and 347 receipts for testing. Each receipt has four entities for extraction: company, address, total, and date. The dataset has relatively complicated and varied layouts and is suitable to validate the generalizability of the model. 

\textbf{CORD} dataset is for both entity extraction and entity linking. It has 800 scanned receipts for the training set, 100 for the validation set, and 100 for the test set. There are in total 4 categories in this dataset, which are further classified into 30 subclasses, such as menu name, total price, etc. We use this dataset to perform entity extraction. 

\textbf{FUNSD} dataset consists of 199 forms annotated with 4 entity types: question, answer, header, and other. It supports both entity linking and entity extraction tasks. The training set has 149 forms, and the test set has 50 forms. We utilize this dataset to evaluate the model's performance on large documents. 

\textbf{Train Tickets} dataset has a total of 2K real documents and 300K synthetic documents. The document image was taken in real-life settings with all the possible conditions, such as dim lighting, distortion, background noise, etc. Entities we need to extract from the train ticket are the ticket number, destination station, seat category, train number, starting station, date, ticket rates, and passenger name. 

Since the dataset does not provide the OCR results, we used the dataset in \citep{DBLP:journals/corr/abs-2004-07464}, which sampled 400 real documents and 1530 synthetic documents from the original datasets and annotated them with bounding boxes and transcripts with OCR. \citep{DBLP:journals/corr/abs-2004-07464} chose 320 real documents and all synthetic documents for training and 80 real documents for testing. The same setting is used by our model for fair comparison. 

\textbf{Business Licenses} dataset contains 320 real documents and 500 synthetic documents. We collect the documents either online or by manually taking the photos in real-life settings. A business license contains nine fields: company name, company type, company start date, registration capital, legal person, operation dates, business scopes, company location, and social credit code. The content consists mainly of numbers and Chinese characters and has different layouts. Since the images are captured in real life, there is inevitable image distortion and background noise. We utilized OCR to extract the transcripts and manually labeled them with different entity types. For synthetic licenses, we first create the templates in variable layouts, then build our corpus for different entity types, and finally synthesize documents~\footnote{Our code for synthesizing the business licenses is publicly available at \url{https://github.com/AYSP/Business-Licenses}.} with the templates and corpus.

\subsection{Experiment Settings}
Our model is implemented in PyTorch and trained with a NVIDIA GTX 3060 GPU with 12GB memory. An Adam optimizer with a decaying learning rate is used. The learning rate is initially set to $1 \mathrm{e}{-4}$ and decays by 0.1 every 50 epochs. The model dimension $d$, the graph module embedding dimension $d^n$ and the BiLSTM hidden dimension $d^b$ are all 512. The sinusoidal embedding dimension $d^b$ is 1024. $K$ is set to 4 for the $Knn$ algorithm in the graph module. The dropout ratio is set to 0.1. Resnet50 \citep{https://doi.org/10.48550/arxiv.1512.03385} with default parameters is used as the image feature extractor. We use the default setting of the transformer encoder in the embedding module.
\subsection{Experiment results}
Since the model is character based, each character in the segment is labeled with the entity type that maximizes the conditional probability of the document series. The label of the segment is decided by the majority of the character labels. For example, the decoded BIO tags of \textbf{01/18} is \textit{B-date}, \textit{I-date}, \textit{O}, \textit{B-date} and \textit{I-date}. While the third character is mislabeled as \textit{O}, the word level prediction is still \textit{date}, which is decided by the majority of the predicted character labels.
\subsubsection{Baseline}
We chose PICK\citep{DBLP:journals/corr/abs-2004-07464} as the baseline method because both PICK and \model{} are graph based, and PICK has been proven effective in document KIE. We compare PICK with our model on datasets including SROIE, CORD, train tickets, and business licenses. We evaluate the model's performance using the entity-level F1 score. 
\subsubsection{Results}
\begin{table}[]
\centering
\resizebox{\columnwidth}{!}{%
\begin{tabular}{|l|cc|cc|cc|}
\hline
\multirow{2}{*}{Dataset} & \multicolumn{2}{c|}{Precision}                 & \multicolumn{2}{c|}{Recall}                    & \multicolumn{2}{c|}{F1}                        \\ \cline{2-7} 
                         & \multicolumn{1}{c|}{Baseline} & GraphRevisedIE & \multicolumn{1}{c|}{Baseline} & GraphRevisedIE & \multicolumn{1}{c|}{Baseline} & GraphRevisedIE \\ \hline
SROIE                    & \multicolumn{1}{c|}{96.79}    & \textbf{96.80}          & \multicolumn{1}{c|}{95.46}    & \textbf{96.04}          & \multicolumn{1}{c|}{96.12}    & \textbf{96.42}          \\ \hline
CORD                     & \multicolumn{1}{c|}{91.75}    & \textbf{93.91}          & \multicolumn{1}{c|}{93.26}    & \textbf{94.61}          & \multicolumn{1}{c|}{92.50}    & \textbf{94.26}          \\ \hline
Train Ticket             & \multicolumn{1}{c|}{98.75}    & \textbf{99.07}          & \multicolumn{1}{c|}{98.45}    & \textbf{98.76}          & \multicolumn{1}{c|}{98.60}    & \textbf{98.91}          \\ \hline
Business License         & \multicolumn{1}{c|}{99.05}    & \textbf{99.37}          & \multicolumn{1}{c|}{99.21}    & \textbf{99.37}          & \multicolumn{1}{c|}{99.13}    & \textbf{99.37}          \\ \hline
\end{tabular}%
}
\caption{Comparison of the baseline model to GraphRevisedIE on four datasets. GraphRevisedIE outperforms the baseline model.}
\label{tab:baseline-comp}
\end{table}
As shown in Table \ref{tab:baseline-comp}, the baseline method achieves competitive performance on the datasets. GraphRevisedIE still outperforms the baseline with small improvements. In comparison to the baseline, the relative positional embedding in GraphRevisedIE is critical in allowing the model to learn the document layout quickly and efficiently. For datasets with relatively fixed layouts, such as train tickets, GraphRevisedIE achieves 0.3\% improvements on the F1 score, although the baseline almost achieves a full score. For the CORD dataset, GraphRevisedIE improves the F1 score by about 1.7\%. Finally, for the SROIE and business license datasets with variable document layouts, GraphRevisedIE still has a 0.2–0.3\% improvement on the F1 score compared to the baseline.

\begin{table}[]
\centering
\resizebox{0.8\textwidth}{!}{%
\begin{tabular}{|c|cc|c|cc|}
\hline
\multirow{2}{*}{Menu} & \multicolumn{2}{c|}{Model}                     & \multirow{2}{*}{Total} & \multicolumn{2}{c|}{Model}                     \\ \cline{2-3} \cline{5-6} 
                      & \multicolumn{1}{c|}{Baseline} & GraphRevisedIE &                        & \multicolumn{1}{c|}{Baseline} & GraphRevisedIE \\ \hline
unitprice             & \multicolumn{1}{c|}{86.96}    & \textbf{96.40}          & emoneyprice            & \multicolumn{1}{c|}{28.57}    & \textbf{40}             \\ \hline
num                   & \multicolumn{1}{c|}{76.19}    & \textbf{95.24}          & total\_price           & \multicolumn{1}{c|}{93.72}    & \textbf{96.15}          \\ \hline
sub\_cnt              & \multicolumn{1}{c|}{90.32}    & \textbf{93.75}          & menutype\_cnt          & \multicolumn{1}{c|}{54.55}    & \textbf{60}             \\ \hline
sub\_price            & \multicolumn{1}{c|}{\textbf{80}}       & 77.78          & menuqty\_cnt           & \multicolumn{1}{c|}{81.97}    & \textbf{84.38}          \\ \hline
discountprice         & \multicolumn{1}{c|}{52.63}    & \textbf{58.82}          & creditcardprice        & \multicolumn{1}{c|}{\textbf{88.89}}    & 85             \\ \hline
\end{tabular}%
}
\caption{Entity level F1 score comparison on the CORD dataset between the baseline method and \model{}. We select the menu and total entity types as examples for easier explanation.}
\label{tab:cord-entity-comp}
\end{table}

Table \ref{tab:cord-entity-comp} illustrates the entity level comparison between the baseline and \model{} on the CORD dataset. Due to the limited space and the large number of different entity types in this dataset, we only selected a subset of entity types for easy elaboration. There are entity types with rich relative positional features. For example, menu.unitprice is usually on the right of menu.nm and is in the rightmost column, while menu.num is usually on the left of menu.nm and is in the leftmost column. The relative positional information can be effectively captured by the relative positional embedding as illustrated in Figure \ref{fig:relative-pe}. GraphRevisedIE outperforms the baseline by a large margin on those entity types. 
\begin{table}[]
\centering
\resizebox{\columnwidth}{!}{%
\begin{tabular}{|l|l|l|l|ccc|ccc|ccc|}
\hline
\multirow{2}{*}{Model} & \multirow{2}{*}{Modality} & \multirow{2}{*}{Pretrained} & \multirow{2}{*}{\# Params} & \multicolumn{3}{c|}{SROIE}                                      & \multicolumn{3}{c|}{CORD}                                       & \multicolumn{3}{c|}{FUNSD}                                      \\ \cline{5-13} 
                       &                           &                              &                          & \multicolumn{1}{c|}{P}     & \multicolumn{1}{c|}{R}     & F     & \multicolumn{1}{c|}{P}     & \multicolumn{1}{c|}{R}     & F     & \multicolumn{1}{c|}{P}     & \multicolumn{1}{c|}{R}     & F     \\ \hline
BERT                   & T                         & \checkmark                   & 340M                     & \multicolumn{1}{c|}{90.99} & \multicolumn{1}{c|}{90.99} & 90.99 & \multicolumn{1}{c|}{88.33} & \multicolumn{1}{c|}{91.07} & 89.68 & \multicolumn{1}{c|}{54.69} & \multicolumn{1}{c|}{67.10} & 60.26 \\ \hline
RoBERTa                & T                         & \checkmark                   & 355M                     & \multicolumn{1}{c|}{91.07} & \multicolumn{1}{c|}{91.07} & 91.07 & \multicolumn{1}{c|}{-}     & \multicolumn{1}{c|}{-}     & -     & \multicolumn{1}{c|}{66.48} & \multicolumn{1}{c|}{66.48} & 66.48 \\ \hline
UniLMv2                & T                         & \checkmark                   & 340M                     & \multicolumn{1}{c|}{94.59} & \multicolumn{1}{c|}{94.59} & 94.59 & \multicolumn{1}{c|}{89.87} & \multicolumn{1}{c|}{91.98} & 90.92 & \multicolumn{1}{c|}{65.61} & \multicolumn{1}{c|}{72.54} & 68.90 \\ \hline
LayoutLM               & T+L                       & \checkmark                   & 343M                     & \multicolumn{1}{c|}{94.38} & \multicolumn{1}{c|}{94.38} & 94.38 & \multicolumn{1}{c|}{94.37} & \multicolumn{1}{c|}{95.08} & 94.72 & \multicolumn{1}{c|}{76.77} & \multicolumn{1}{c|}{81.95} & 79.27 \\ \hline
LayoutLMv2             & T+L+V                     & \checkmark                   & 426M                     & \multicolumn{1}{c|}{96.25} & \multicolumn{1}{c|}{96.25} & 96.25 & \multicolumn{1}{c|}{94.53} & \multicolumn{1}{c|}{95.39} & 94.95 & \multicolumn{1}{c|}{80.29} & \multicolumn{1}{c|}{85.39} & 82.76 \\ \hline
PICK                   & T+L+V                     & \xmark                       & -                        & \multicolumn{1}{c|}{96.79} & \multicolumn{1}{c|}{95.46} & 96.12 & \multicolumn{1}{c|}{91.75} & \multicolumn{1}{c|}{93.26} & 92.50 & \multicolumn{1}{c|}{-}     & \multicolumn{1}{c|}{-}     & -     \\ \hline
GraphRevisedIE         & T+L+V                     & \xmark                       & 68M                      & \multicolumn{1}{c|}{96.80} & \multicolumn{1}{c|}{96.04} & 96.42 & \multicolumn{1}{c|}{93.91} & \multicolumn{1}{c|}{94.61} & 94.26 & \multicolumn{1}{c|}{76.67} & \multicolumn{1}{c|}{80.22} & 78.41 \\ \hline
\end{tabular}%
}
\caption{We compare the model's performance with other models, including the large version of the pretrained models and PICK.}
\label{tab:performance}
\end{table}

\begin{figure*}
    \centering
    \includegraphics[scale=0.25]{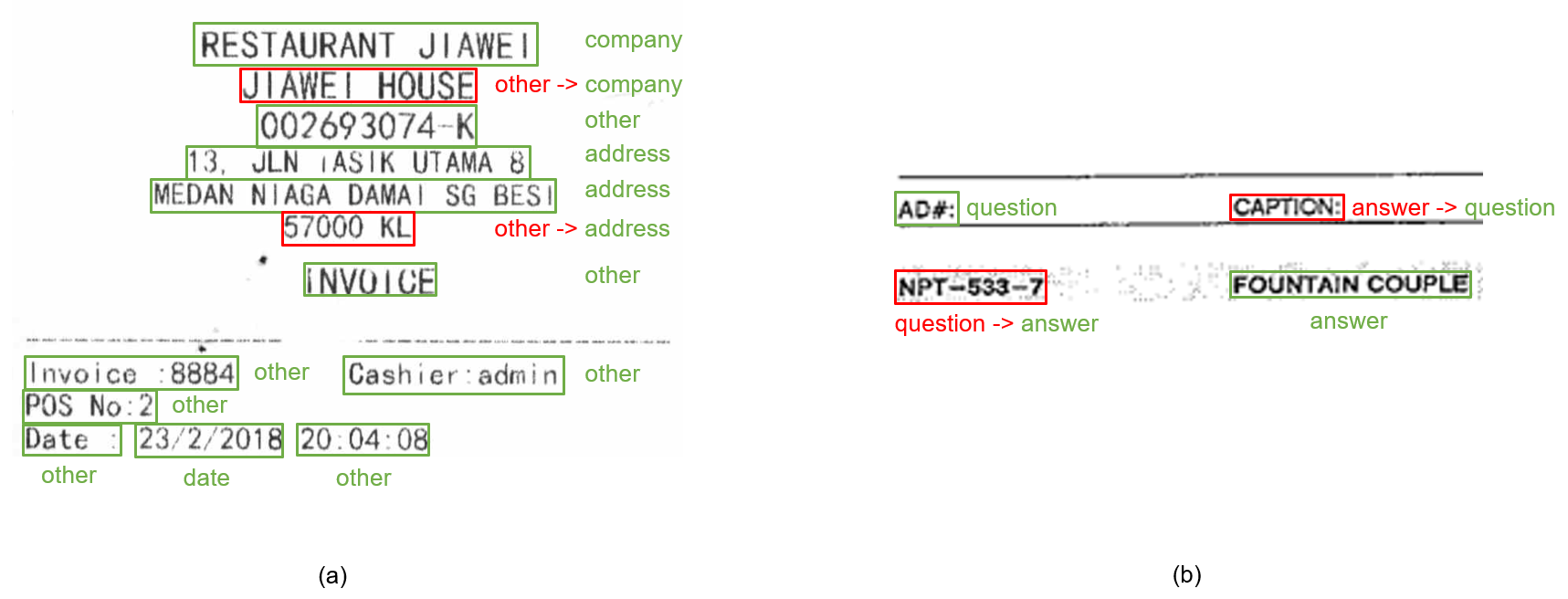}
    \caption{Example predictions made by \model{} on (a) SROIE receipt and (b) FUNSD form. (a) \textbf{JIAWEI HOUSE} is incorrectly labeled as \textbf{other} when it should be part of the company, and \textbf{57000 KL} is incorrectly labeled as \textbf{other} when it should be part of the address. (b) The question answer pairs should be in the column direction, while \model{} bases the prediction on the row direction. The performance of \model{} can be further improved by using more powerful textual embedding to deal with isolated semantic context.}
    \label{fig:example-predictions}
\end{figure*}
Table \ref{tab:performance} presents the model performance on the public SROIE, CORD, and FUNSD datasets. Performance metrics of the pretrained models are obtained from their original papers. Particularly, we compare our model with the large version of the pretrained models. On the SROIE dataset, GraphRevisedIE achieves the highest F1 score even with fewer parameters and without pre-training. On the CORD dataset, \model{} achieves comparable performance with the pretrained models, which proves its generalization ability on small datasets with varied layouts. Since the CORD dataset has many more entity labels than SROIE, we speculate that richer textual embedding can help the model learn the label semantics better and reduce the ambiguity. Therefore, pre-training methods achieve the best performance. Despite its effectiveness on small documents, GraphRevisedIE does not perform well on large and complicated documents such as the FUNSD form. A large document usually contains richer semantic information than a small document. Since our model only leverages the vanilla one-hot textual embedding, it is hard to embed the semantic features effectively, especially when the semantic context is split into multiple segments as shown in Figure \ref{fig:example-predictions}. This observation is aligned with the findings in the CORD dataset. Other limitations of GraphRevisedIE include the requirement of some initial experiments to determine the optimal K for the graph module and post-processing to generate the word level predictions since our model is character based. 

\subsubsection{Ablation Study}
\label{sec:ab}

\begin{table}[]
\centering
\Scale[0.6]{
\resizebox{\columnwidth}{!}{
\begin{tabular}{|l|l|l|}
\hline
& Train Tickets & Business Licenses \\ \hline
Full Model                   & 98.9                  & 99.3        \\ \hline
w/o spatial features          &  $\downarrow$0.6             & $\downarrow$0.5           \\ \hline
w/o textual features          & $\downarrow$0.7            & $\downarrow$5.4           \\ \hline
w/o image features          & $\downarrow$0.3              & $\downarrow$0.2           \\ \hline
w/o graph module           & $\downarrow$0.4             & $\downarrow$0.7           \\ \hline
\end{tabular}
}}
\caption{Ablation study on train tickets and business licenses to evaluate the importance of each component in the framework.}
\label{tab:features-ab}
\end{table}

In this section, we first do an ablation study on the train ticket and business license datasets to evaluate the importance of components in the model. As is shown in Table \ref{tab:features-ab}, one observation is that removing the textual feature does not reduce the F1 score much on the train ticket dataset. The model can still rely on the layout and visual features to achieve good performance. For the business licenses dataset, the F1 score decreases by 5.4\% when the textual feature is removed. This is aligned with the fact that the business license has more semantic information than the train ticket. Intuitively, visual features also contribute to the model's performance. Information such as font, color, size, background, etc. can only be captured by the visual encoder, which is important to reduce ambiguity. Last but not least, since the graph module enriches the character embedding with global context, removing the graph module causes decreased F1 scores on both datasets. 

\begin{figure*}
    \centering
    \includegraphics[scale=0.6]{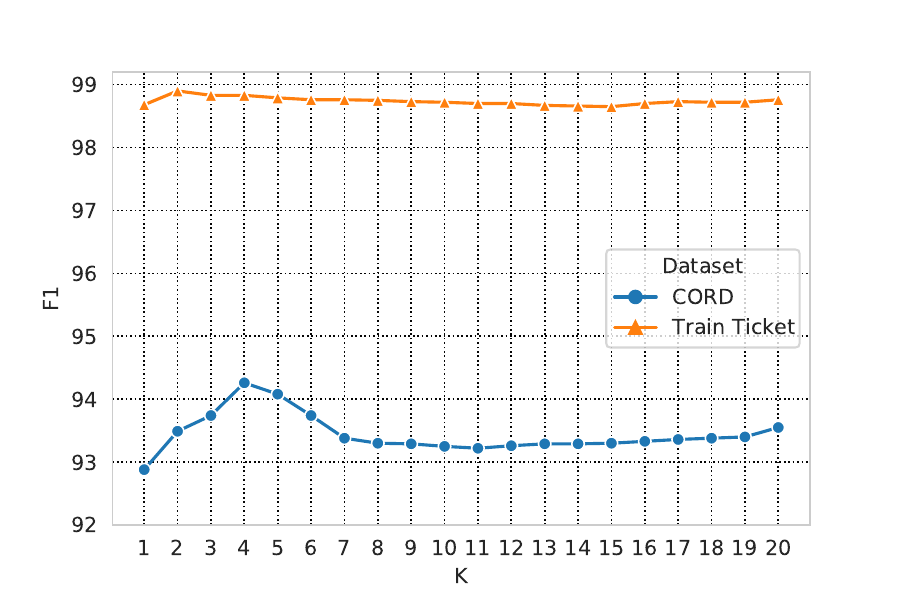}
    \caption{Experiments on the CORD and Train Ticket datasets to evaluate model performance under different $K$ in the $Knn$ algorithm.}
    \label{fig:k-ab}
\end{figure*}

We perform another ablation study to evaluate the impact of $K$, the number of neighbors used in the $Knn$ algorithm in the graph module, on the CORD and train ticket datasets. From Figure \ref{fig:k-ab}, gradually increasing $K$ from 1 results in better performance, and then the performance starts to decrease after some point. For the CORD dataset, the best F1 score is achieved when $K$ equals 4. Further increasing $K$ only brings in more noise from distant segments and reduces the performance. For the train tickets, similarly, the best F1 score is achieved when K is 2. This study demonstrates that a sparse graph, i.e., K is small, better captures the document graph than a dense graph, i.e., K is large, on the CORD and train ticket datasets and results in optimal performance. 

\section{Conclusion}
In this paper, we propose a novel framework named \model{}, which can effectively combine multimodal features from VRD, to perform the KIE task. We integrate with a graph module to model the underlying document graph, which is used to propagate the global context among segments to enrich the character embedding. \model{} is has been proven to achieve good performance on multiple public datasets, and it is able to generalize over small documents with varied layouts. It's worth mentioning \model{} does not perform well on large documents. Replacing character level embedding with pretrained word level embedding to utilize more semantic features can possibly improve the model's performance. Besides, although \model{} provides flexibility to customize the number of neighbors in the graph module for different datasets, it adds more manual effort and could be automated. Finally, we set the kernel function to be the dot product in the graph module, while more options could be explored in future work. 
\label{}




 \bibliographystyle{elsarticle-num-names} 
 \bibliography{GraphRevisedIE.bib}

\end{document}